\documentclass[twocolumn,showpacs,preprintnumbers,amsmath,amssymb,prb,superscriptaddress]{revtex4}

\usepackage{graphicx}% Include figure files
\usepackage{latexsym}
\usepackage{dcolumn}% Align table columns on decimal point
\usepackage{bm}% bold math

\begin{document}

\title{ Numerical study of hydrogenic effective mass theory for an \\ impurity P donor in Si in the presence of an electric field and interfaces }

\author{L.M. Kettle \email{s335321@student.uq.edu.au}}
\affiliation{Centre for Quantum Computer Technology.}
\affiliation{Centre for Computational Molecular Science.}
\affiliation{University of Queensland, Brisbane Queensland 4072 Australia.}
\author{H.-S. Goan}
\affiliation{Centre for Quantum Computer Technology.}
\affiliation{School of Physics, University of New South Wales, Sydney New South Wales 2052 Australia.}
\author{Sean C. Smith}
\affiliation{Centre for Computational Molecular Science.}
\affiliation{University of Queensland, Brisbane Queensland 4072 Australia.}
\author{C.J. Wellard}
\affiliation{Centre for Quantum Computer Technology.}
 \affiliation{School of Physics, University of Melbourne, Melbourne Victoria 3010 Australia.}
\author{L.C.L. Hollenberg}
\affiliation{Centre for Quantum Computer Technology.}
 \affiliation{School of Physics, University of Melbourne, Melbourne Victoria 3010 Australia.}
\author{C.I. Pakes}
\affiliation{Centre for Quantum Computer Technology.}
 \affiliation{School of Physics, University of Melbourne, Melbourne Victoria 3010 Australia.}

\date{\today}

\begin{abstract}
In this paper we examine the effects of varying several experimental parameters in the Kane quantum computer architecture: $A$-gate voltage, the qubit depth below the silicon oxide barrier, and the back gate depth to explore how these variables affect the electron density of the donor electron. In particular, we calculate the resonance frequency of the donor nuclei as a function of these parameters. To do this we calculated the donor electron wave function variationally using an effective mass Hamiltonian approach, using a basis of deformed hydrogenic orbitals. This approach was then extended to include the electric field Hamiltonian and the silicon host geometry. We found that the phosphorous donor electron wave function was very sensitive to all the experimental variables studied in our work, and thus to optimise the operation of these devices it is necessary to control all parameters varied in this paper.
\end{abstract}


\maketitle

\section{\label{sec:one} Introduction}

Since Kohn and Luttinger's\cite{kohn1,kohn2} original work on shallow donors in silicon, there has been renewed interest in the study of donor impurities in silicon, particularly the Si:$^{31}\mbox{P}$ system, following Kane's\cite{kane} proposal for a solid-state quantum computer. In the Kane quantum computer, information is encoded onto the nuclear spins of donor phosphorous atoms in doped silicon electronic devices. Application of an electrostatic potential at surface electrodes positioned above the qubits ($A$-Gates) tunes the resonance frequency of individual spins, while surface electrodes between qubits ($J$-Gates) induces electron-mediated coupling between nuclear spins. Perturbing the donor electron density with an externally applied electric field is crucial in tuning the hyperfine interaction between the donor electron and nucleus and hence also in tuning the resonance frequency of the P nuclei and controlling logical operations. Substantial theoretical efforts have been devoted to modeling the P donor electron ground state in the silicon wafer device, and the altered ground state with an externally applied electric field. In this paper we discuss relevant experimental parameters which can be controlled to perturb the donor electron wave function.

There is a considerable amount of work done in this area, and several theoretical approaches have been pursued with varying degrees of application and approximation. In Kohn and Luttinger's\cite{kohn1,kohn2} work, the P donor ground state in the bulk silicon is calculated using a single trial wave function: a deformed $1S$ hydrogenic orbital and varying the Bohr radii to minimise the ground state energy. In this paper we follow Faulkner's\cite{faulkner} approach and extend Kohn's method to include a trial wave function expanded in a basis of deformed hydrogenic orbitals, and vary the Bohr radii to minimise the ground state energy. As we have used a large basis set in this approach, the ground state wave function has the flexibility to distort with the application of an electric field above the P donor. Several authors\cite{koiller,tahan,sham} have previously investigated the effects induced by strain and interface regions on donor states. These external influences partially lift the valley degeneracy in the bulk silicon.

The effect of an electric field potential at a gate above a P donor in a silicon substrate on the hyperfine interaction coupling between the P donor electron and nucleus has already been reported by several authors. In the work of Kane\cite{kane2} and Larinov \emph{et al.},\cite{larionov} the effect of an electric field potential in the bulk silicon host is considered using perturbative theory, excluding the additional interface potentials. Wellard \emph{et al.}\cite{lloyd} consider both the influence of the electric field and interface barriers using a spherical effective mass Hamiltonian.

The main advantage demonstrated in our approach using the anisotropic basis is the flexibility in choosing the smaller effective Bohr radius for the donor ground state to be in the direction towards the interface regions. This minimizes the overlap of the donor wave function into these regions. For shallow donor depths, the donor wave function is restricted in moving towards the $A$-gate because of the silicon oxide interface. 

In this work, we include the effects of both the electric field potential and the interface regions, and the anisotropy of the conduction band minimum in Si. To our knowledge, there have been hitherto no published results for modeling electrostatic gate operations in the Kane quantum computer which include simultaneously the anisotropy of the effective masses in the silicon host, the electric field potential and the interface regions in the Si wafer device. In this paper we address all these criteria and discuss relevant experimental parameters which can be adjusted to fine tune the contact hyperfine interaction. We calculate this coupling as a function of $A$-gate voltage, donor depth below associated $A$-gate and the back gate depth. A subsequent paper will discuss our further results for the $J$-gate controlled electron exchange interaction between adjacent donor electrons.

In Sec.~\ref{sec:two}, we will discuss some background effective mass theory and the approach we took to obtain the phosphorous donor ground state in bulk silicon with no electric field applied. Section~\ref{sec:three} discusses how we obtained the electric field potential and modeled the silicon host geometry to include the silicon oxide layer and back gate. The numerical results using the methods outlined in the previous sections are presented in Sec.~\ref{sec:four} for the varying experimental parameters studied. Finally we summarise our major findings in Sec.~\ref{sec:five}.

\section{\label{sec:two} Faulkner's Method}

Neglecting inter-valley terms, the one-valley effective mass equation for the energy levels of donors in silicon is given below:\cite{faulkner}

\begin{eqnarray}
-\bigg{\lbrack} \frac{\hbar^2}{2m_{\bot}}\left( \frac{\partial^2}{\partial x^2} + \frac{\partial^2}{\partial y^2} \right) + \frac{\hbar^2}{2m_{\parallel}} \frac{\partial^2}{\partial z^2} + \frac{e^2}{\epsilon r} \bigg{\rbrack} \Psi(r) &=& E \Psi(r), \nonumber \\ && \label{a1}
\end{eqnarray}
where $\epsilon = 11.4$ is the dielectric constant, and $m_{\bot} = 0.1905 m_0$ and $m_{\parallel}= 0.9163 m_0$ are the transverse and longitudinal effective masses respectively, and $m_0$ is the mass of a free electron. Here we are expanding the energy, $E_k^0$ around the conduction band minimum along the $z$-axis at $\vec{k} = (0,0,k_0)$:

\begin{eqnarray}
E_k^0& =& E_0^0 + \frac{\hbar^2}{2m_{\bot}} \left( k_x^2 + k_y^2 \right) + \frac{\hbar^2}{2m_{\parallel}} \left( k_z^2 - k_0^2\right) .
\end{eqnarray}

We followed Faulkner's approach and kept the full anisotropy of the conduction band minimum. We expanded the donor electron wave function, $\Psi(r)$ in a basis of deformed hydrogenic orbitals:
\begin{eqnarray*}
\Psi(r)& =& \sum_{nlm} \left( \frac{\beta}{\gamma}\right)^{1/4} \psi_{nlm}(x,y, \sqrt{ \frac{\beta}{\gamma} } z,a),
\end{eqnarray*}
where $\psi_{nlm}(x,y,z,a) = R_{nl}(a,r) Y_{lm}(\theta, \phi)$ are the normalised hydrogenic orbitals, $\gamma =  m_{\perp}/m_{\parallel} = 0.2079$, $a$ is the effective Bohr radius in the $x,y$ directions, and $\beta$ is an adjustable parameter which gives the effective Bohr radius $b$ in the $z$ direction.

If we use atomic units, where the unit of length $ a_b =  \hbar^2 \epsilon / m_\perp e^2= 31.7 ${\AA} and unit of energy  $ m_{\perp} e^4 / 2 \hbar^{2} \epsilon^2  = 19.94$meV, Eq.~(\ref{a1}) becomes:
\begin{eqnarray}
-\left( \frac{\partial^2}{\partial x^2} + \frac{\partial^2}{\partial y^2} + \gamma \frac { \partial^2 }{ \partial z^2 }  + \frac {2}{r} \right) \Psi(r) &=& E \Psi(r), \nonumber \\ && \label{a2}
\end{eqnarray}

\begin{figure}
\includegraphics[height=3in,width=2.5in,angle=-90]{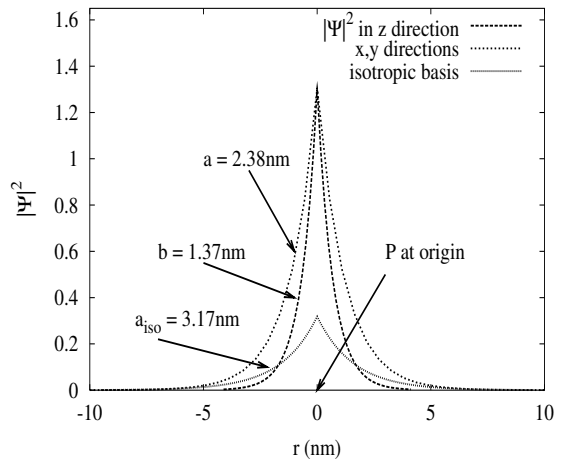} %,width=8cm,height=13cm,clip=,angle=270}
\caption{\label{fige6} Ground state electron density without electric field. }
\end{figure}

Equation~(\ref{a2}) was diagonalised with the effective Bohr radius $a$ and parameter $\beta$ varied to minimise the ground state energy $E$. The ground state energy converged using a basis of 91 hydrogenic orbitals to give $E = -31.23\mbox{meV}$, and effective Bohr radii: $ a = 23.81${\AA}  and $ b = \sqrt{(\gamma / \beta)}  a =13.68 ${\AA}. These results are consistent with  Kohn's results of  $ a = 25 ${\AA}  and $b = 14.2 $,{\AA}\cite{kohn1,kohn2} and Faulkner's ground state energy $E = -31.27\mbox{meV}$\cite{faulkner} for phosphorous. 

The ground state wave function obtained was a deformed hydrogenic $1S$ orbital. Figure~\ref{fige6} shows the ground state electron density plotted in the $x,y$  and $z$ directions for comparison of the different effective Bohr radii obtained in the different directions. Also shown in this figure is the ground state obtained using a spherical effective mass Hamiltonian and isotropic hydrogenic orbitals as a basis, here the effective electron mass is given by $m_* \approx m_{\bot} = 0.1905 m_0 $, which gives an effective Bohr radius of 3.17nm.

\section{\label{sec:three} Including the electric field and silicon host potential}

Faulkner's method was then extended to include the effects of an electric field above the qubit, and boundary conditions of the silicon host. The solution of Poisson's equation to extract the electric field potential for our device with the $A$-gate at varying voltages was obtained by simulation using a Technology Computer Aided Design (TCAD) modeling package.\footnote{Technology Computer Aided Design modeling package, Integrated Systems Engineering AG, Zurich}

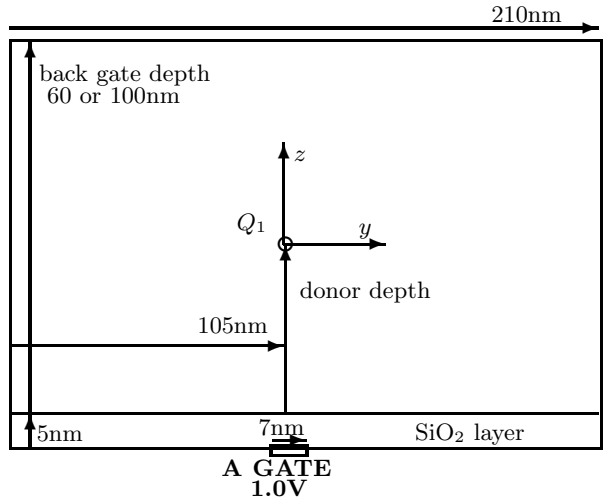
\begin{figure}

\setlength{\unitlength}{0.09cm}
\begin{picture}(100,70)
\thicklines
\put(6,7){ \framebox(87,60) {}}

\put(7,12){\line(1,0){87}}

\put(7,69){\vector(1,0){87}}
\put(84,71){\makebox(0,0) {210nm   }}

\put (7,22){\vector(1,0){40.5}}
\put(40,25){\makebox(0,0) {105nm}}

\put(47.5,37){\vector(1,0){15}}

\put(47.5,37){\vector(0,1) {15}}

\put(50.5,50){\makebox(0,0) {$z$   }}

\put(59.5,39){\makebox(0,0){$y$}}

\put(10,7) {\vector(0,1){5}}

\put(75,9) {\makebox(0,0){ $\mbox{SiO}_2$ layer }}
\put(15,9) {\makebox(0,0){5nm }}

 \put(45.75,8) {\vector(1,0){5}}
 \put(47.75,10) {\makebox(0,0){7nm }}

\put(10,12) {\vector(0,1){55}}

\put(24,62) {\makebox(0,0){ back gate depth }}
\put(23,59) {\makebox(0,0){60 or 100nm }}

\put(47.75,37){\circle{2}}

\put(43.5,40){\makebox(0,0){\bfseries {$Q_1$}  }}

\put(47.75,12) {\vector(0,1){25}}

\put(59.5,30) {\makebox(0,0){donor depth}}

\put(45.75,6){\framebox(5,1){}       }

\put(46.5,4){\makebox(0,0){ \bfseries {A GATE}  }            }

\put(47.5,1){\makebox(0,0){\bfseries{1.0V} }}

\end{picture}

\caption{\label{fig:fige1} Schematic design parameters implemented in TCAD to model the Kane computer architecture.}

\end{figure}

TCAD is used in the electronics industry as a tool for 2-D and 3-D modeling and simulation of semiconductor devices. It employs a coupled Newton-like solver at discrete nodes to obtain the self-consistent solution of the Poisson and electron-hole continuity equations. Figure~\ref{fig:fige1} shows the 2-D device scheme implemented in TCAD used to model the application of a voltage to the $A$-gate above qubit, $Q_1$. The lateral edges of the silicon lattice were assumed to extend infinitely in the $y$-direction, but the electrostatic potential was only obtained on a finite grid 210nm wide, with the potential set to zero outside this region. We checked that this approximation is valid at the boundaries and found the TCAD potential had fallen close to zero ($10^{-4} - 10^{-5}$eV), at $y = \pm 105$nm. The potential in 2-D from TCAD is assumed to have a ``thickness'' in the third dimension ($x$) of 1$\mu$m. 

In this paper we examine the effects of varying several experimental parameters: $A$-gate voltage, qubit depth below the silicon oxide barrier, and the back gate depth to explore how these variables affect the electron density of the donor electron at the phosphorous nuclei. In particular we calculate the resonance frequency of the donor nucleus as a function of these parameters.

The application of a potential, and the silicon host geometry in the device shown in Fig.~\ref{fig:fige1} splits  the degeneracy of the two local minima along the $z$-axis, compared to the other four along the $x$ and $y$ axis in the lower $A_1$ ground state.\cite{koiller} With no electric field applied the ground state wave function is $> 99\%$ $1S$ in character. When the voltage applied is low enough so that the wavefunction stays predominantly $1S$ in character, diagonalising the single valley effective mass equation is equivalent for solving in either valley, $\pm z$, since the deformed $1S$ wavefunction is symmetric in $z$.\cite{tahan}

Using these justifications we can formulate the problem using a co-ordinate system with the $z$-axis in the direction from $Q_1$ to the interface. With this convention we expand the donor wave function around the conduction band minimum oriented along the $z$-axis. Because of the smaller effective Bohr radius in the  $z$ direction towards the interface and back gate, the ground state is lower in energy since there is less penetration of the wave function into these barrier regions. 

With the electric field the Hamiltonian is: $H = H_0 + H_1$, where $H_0$ is the zero field Hamiltonian, and $H_1=  V(y,z)$ is the electric field potential term. $V(y,z)$ is the electric field potential generated from TCAD, and here we also add an additional term to model the  $\mbox{SiO}_2$ layer and the back gate. The  $\mbox{Si}/\mbox{SiO}_2$ barrier was modeled as a step function with height 3.25eV, since most insulators have a work function greater than $ 3$eV.\cite{Yu} The back gate serves as a reference voltage point (ground) to the voltages applied to the top gates. Outside the back gate the potential was set at 3.25eV also.

To calculate the perturbed donor electron wave function and energies we constructed the electric field Hamiltonian matrix, $H_1$, with its elements given by:

\begin{eqnarray}
\lefteqn{ \langle n'l'm' \vert H_1 \vert nlm \rangle }  \nonumber \\
&=& \sqrt{ \frac{\beta}{\gamma} }\int dx^3 \psi^*_{n'l'm'}(x,y, \sqrt{ \frac{\beta}{\gamma} } z,a) V(y,z) \nonumber \\ 
&& \times \psi_{nlm}(x,y, \sqrt{ \frac{\beta}{\gamma} } z,a).  \label{ef1}
\end{eqnarray}

The integrals in Eq.~(\ref{ef1}) were then calculated numerically for the varying voltages at the $A$-gate and qubit position. Once $H_1$ was obtained the total Hamiltonian was then diagonalised to find the donor electron ground state with the varying experimental parameters.

\section{\label{sec:four} Numerical Results}

\subsection{ Results obtained varying $A$-gate voltage and donor depth}

 \begin{figure} 
 \includegraphics[height=3in,width=2.5in,angle=-90]{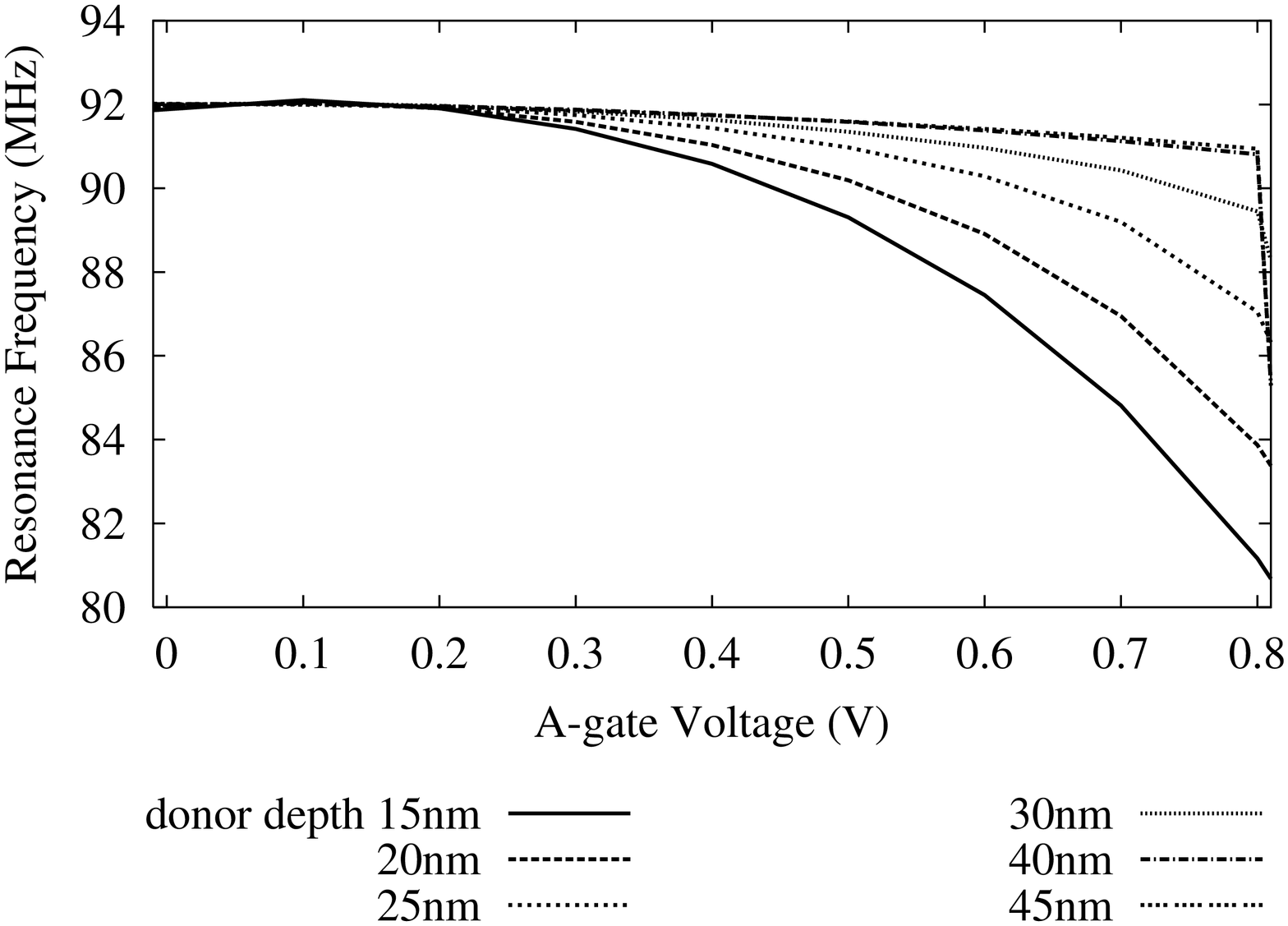} %,width=7cm,height=11cm,clip=,angle=270}
 \caption{\label{fig:fige3} Nuclear resonant frequency shifts of qubit $Q_1$ at lower voltages with varying donor depths, back gate depth at 60nm, using anisotropic basis. }
% \end{figure}
% \begin{figure} 
 \includegraphics[height=3in,width=2.5in,angle=-90]{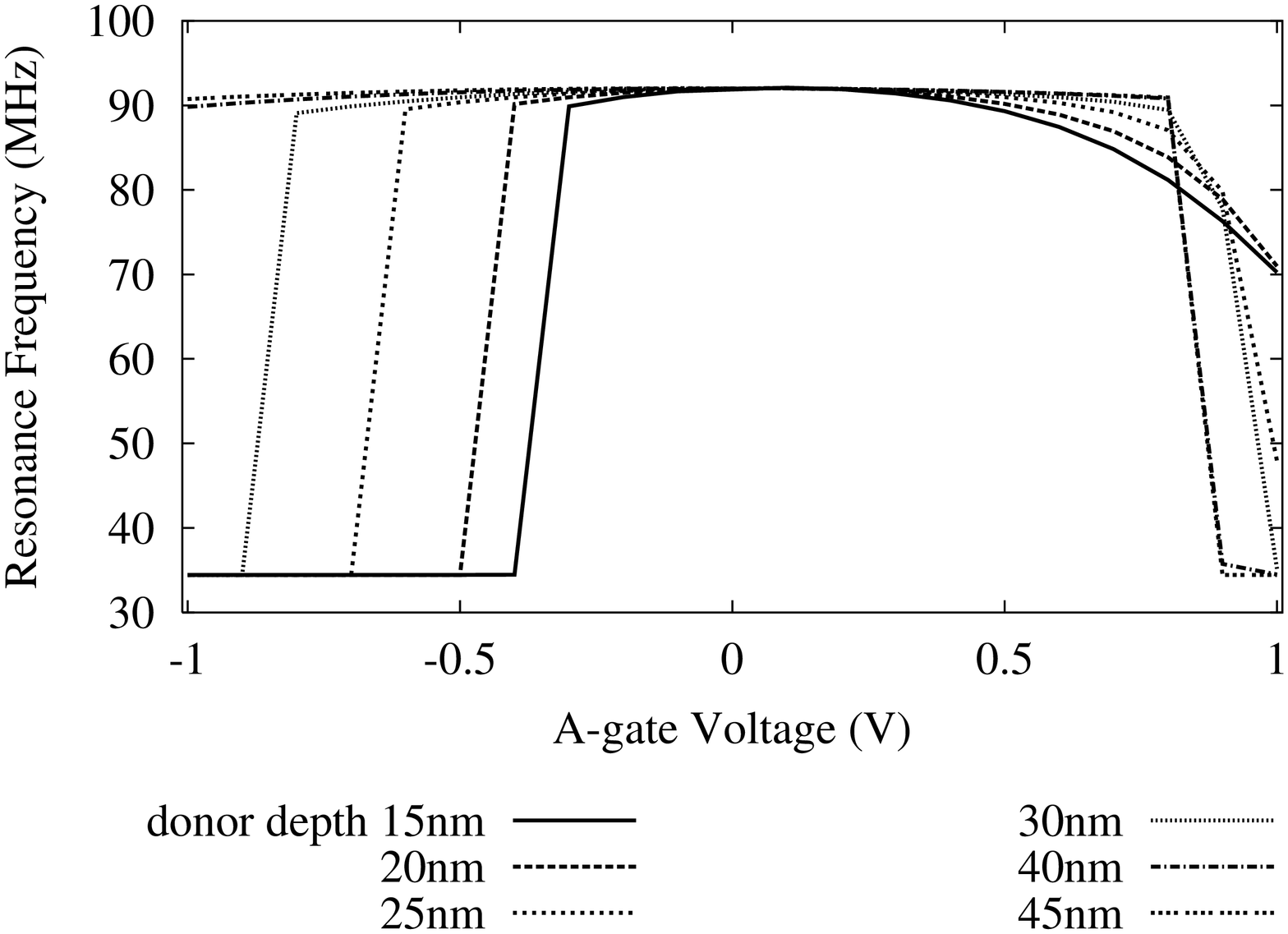}  % ,width=7cm,height=11cm,clip=,angle=270}
 \caption{\label{fig:fige4} Nuclear resonant frequency shifts of qubit $Q_1$ at varying voltage and donor depth, back gate depth at 60nm, using anisotropic basis. }
\end{figure}

The perturbed donor electron ground state was calculated for each set of experimental parameters to compare and optimise the conditions for addressing the target qubit, $Q_1$. Once the electron ground state was found we calculated the value of the contact hyperfine coupling $A(V)$ for each particular voltage at the $A$-gate and qubit depth below this gate. 

The general formula for the contact hyperfine coupling $A(V)$ is given below:
\begin{eqnarray}
A(V) &=& \frac{2}{3} \mu_B g_N \mu_N \mu_0  |\Psi (V,0)|^2,
\end{eqnarray}
where $\Psi (V,0)$ is the donor electron ground state wave function evaluated at the donor nucleus, $\mu_B$ is the Bohr magneton, $g_N$ is Lande's factor for $^{31}$P,  $\mu_N$ is the nuclear magneton and $\mu_0$ is the permeability of free space.\cite{larionov,lloyd}

Since we use effective mass theory, instead of calculating the donor wave function with the full expansion of the Bloch functions, we calculate the envelope function, which describes the smooth donor-related modulation of the electron wave-function. So instead of calculating the contact hyperfine coupling, $A(V)$, directly we calculate the relative shift in $A(V)$ with the potential applied and assume this shift will be similar to those of the true wave function.\cite{lloyd} Thus we need to calculate:
\begin{eqnarray}
A(V) &=& \frac{|\Psi(V,0)| ^2}{|\Psi(0,0)|^2 }A(0),
\end{eqnarray}
where $A(0)/h = 28.76$MHz is determined for $^{31}\mbox{P}$ in silicon from experimental data,\cite{larionov,kane} and $\Psi(V,r)$ are the donor envelope wave functions calculated by our method.

The phosphorous nuclear resonant frequency is affected by the donor electron when the valence electron is spin polarized by a background magnetic field, $B$, of the order of 2T. The hyperfine interaction constant is related to the frequency separation of the nuclear levels, via the following equation (accurate to second order):\cite{kane}
\begin{eqnarray}
h \nu &=& 2 g_N \mu_N B + 2A + \frac{2A^2}{\mu_B B}. 
\end{eqnarray}

In all the calculations we considered the background magnetic field fixed at 2T. Figure~\ref{fig:fige3} shows the nuclear resonant frequency shift of $Q_1$, calculated for a lower range of positive $A$-gate voltages, between 0V and 0.8V, for the varying donor depths below the silicon oxide barrier. Figure~\ref{fig:fige4} shows the nuclear resonant frequency shift calculated for the full range of $A$-gate voltages, between $-1.0$V and $1.0$V, for the varying donor depths below the silicon oxide barrier. These plots are calculated with a close back gate depth set at 60nm

For comparison of our method with previous results~\cite{lloyd} reported using a spherical effective mass Hamiltonian, we calculated the resonance frequency of $Q_1$ using an isotropic effective Bohr radius of $\approx$ 3nm. Our results were consistent with the calculations of Wellard \emph{et al.}~\cite{lloyd} The results for the isotropic basis showed that for donor depths close to the silicon oxide barrier, the wave function was restricted in moving towards the applied $A$-gate voltage. The donor wave function obtained using the anisotropic basis, is advantageous because of the smaller effective Bohr radius in the direction toward the silicon oxide layer, which results in less penetration of the donor wave function into the interface regions. Thus the anisotropic basis produced a more energetically favorable ground state than the isotropic ground state. 

\begin{figure}
 \includegraphics[height=3in,width=2.5in,angle=-90]{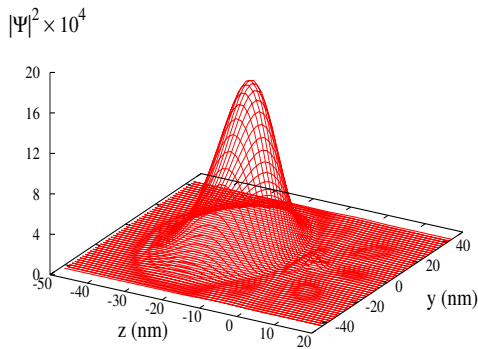} % ,width=7cm,height=11cm,clip=,angle=270}
 \caption{\label{fig:fige7a}  Ground state electron density in $yz$-plane for donor depth at 40nm and voltage at 1V at the $A$-gate. }
 \end{figure}

For the lower voltages ($\leq 0.8$V), the results are consistent with the expectation that the closer the donor depths are to the applied voltage, the greater the frequency shift. At voltages above a certain threshold and donor depths further away from the silicon oxide barrier, there is a huge difference in the donor wave function from the zero field ground state, as it is perturbed almost completely away from the nucleus. Figure~\ref{fig:fige7a} shows an example of this change in electron density for a voltage of 1.0V at the $A$-gate and donor depth of 40nm. Here the P nucleus is at the origin and as $z$ decreases the electric field increases.

\begin{figure} 
 \includegraphics[height=3in,width=2.5in,angle=-90]{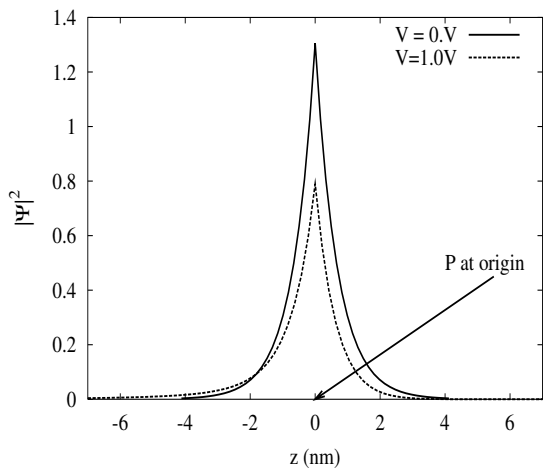} % ,width=7cm,height=11cm,clip=,angle=270}
 \caption{\label{fig:fige7}  Ground state electron density for donor depth at 20nm and voltage at 0. and 1.0V at the $A$-gate, in the $z$-direction. }
\end{figure}

\begin{figure} 
 \includegraphics[height=3in,width=2.5in,angle=-90]{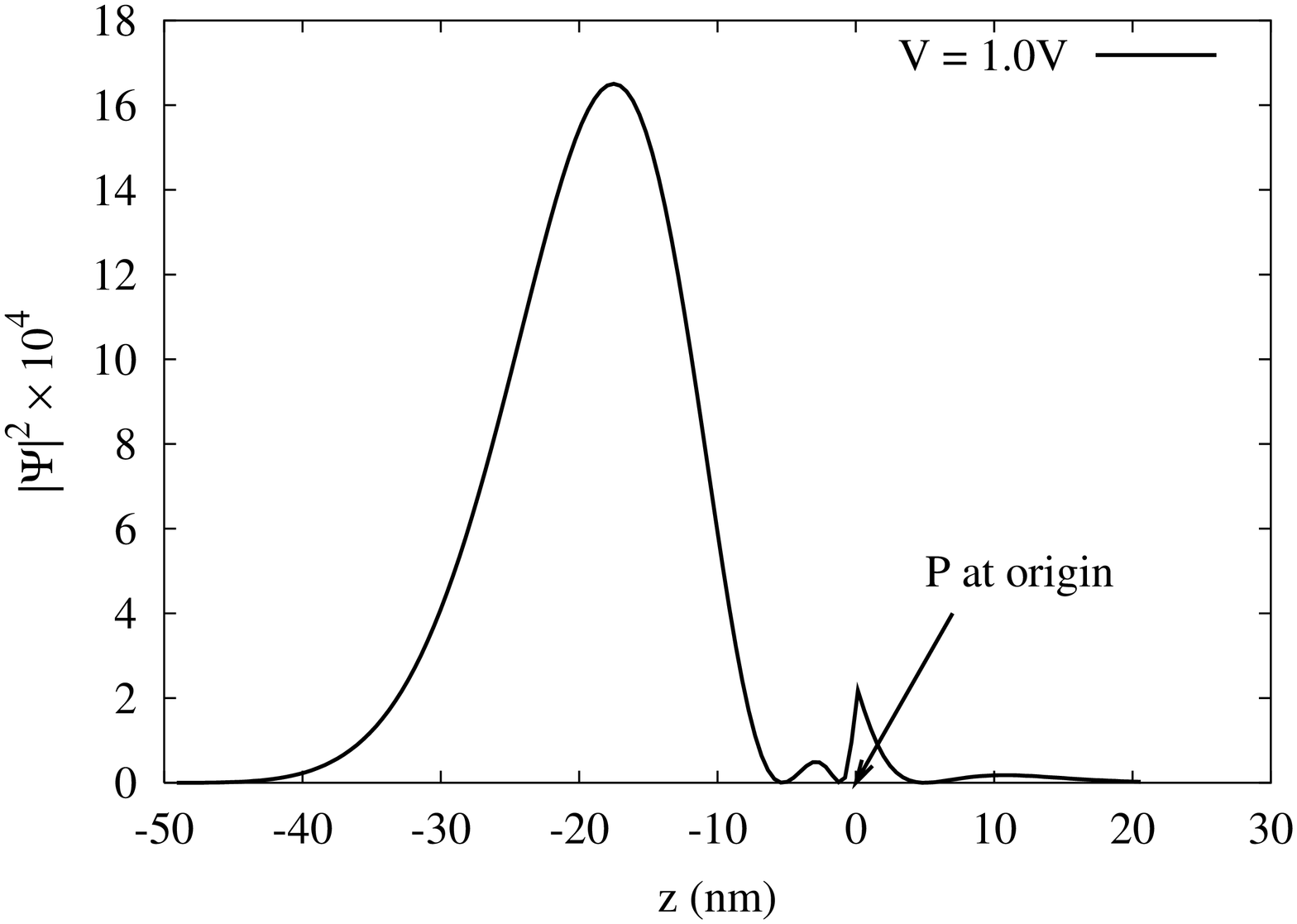}  %,width=7cm,height=11cm,clip=,angle=270}
 \caption{\label{fig:fige8}  Ground state electron density for donor depth at 40nm and voltage at 1.0V at the $A$-gate, in the $z$-direction. }
 \end{figure}

In Fig.~\ref{fig:fige7} and~\ref{fig:fige8} we observe the difference in the donor electron ground state obtained for a donor depth of 20 and 40nm with a positive voltage of 1.0V at the $A$-gate. In both these plots the donor wave function moves toward the applied $A$-gate voltage in the negative $z$ direction. For a close donor depth of 20nm we observe that even though the donor wave function moves slightly toward the $A$-gate, it is significantly restricted in moving in this direction because of the silicon oxide interface in this direction also. In contrast the donor wave function for a depth of 40nm deforms unhindered toward the $A$-gate, and most of the electron density has been transformed away from the nucleus. 

 \begin{figure}
  \includegraphics[height=3in,width=2.5in,angle=-90]{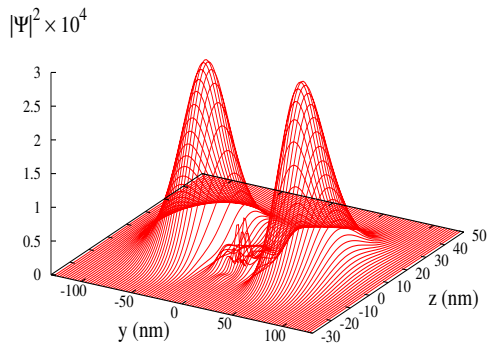} %,width=7cm,height=11cm,clip=,angle=270}
 \caption{\label{fig:fige8b}  Ground state electron density in $yz$-plane for donor depth at 20nm and voltage at -1.0V at the $A$-gate. }
 \end{figure}

Figure~\ref{fig:fige8b} shows the donor electron density obtained in the $yz$-plane for a negative voltage of -1.0V at the $A$-gate and a donor depth of 20nm. A negative applied voltage causes the electron to disperse in all directions away from the positive potential, this plot demonstrates that because of the close back gate in the positive $z$-direction, the electron density predominantly perturbs away from the applied voltage in either direction laterally. 

Because of the interface regions it is either energetically favorable for the donor electron wave function at shallow donor depths to distort completely away from the nucleus, when the gate voltage is negative, or for the donor wave function to be restricted in distorting towards the $A$-gate, with a positive voltage. 

In Table~\ref{table1} we present a comparison of the difference in the ground state energy for the donor wave function without the electric field ($E_0$), and with a positive voltage of 1.0V applied to the $A$-gate ($E_{1V}$). Also reported in this table is the TCAD potential at the P nucleus for the varying donor depths.
\begin{center}
\begin{table}  [h!]
\caption{$ E_{1V} -E_{0}$ for a back gate depth of 60nm. \label{table1} }
\vspace{0.3cm}
\begin{tabular} {|l|l|l|} \hline
$Q_1$ Depth & TCAD Potential & $ E_{1V} -E_{0}$   \\
 & at $Q_1$ (meV) &  (meV)  \\  \hline
20nm & -90.02   &  -91.70  \\ \hline
40nm & -37.06 & -47.73  \\ \hline
\end{tabular}
\end{table}
\end{center}

For the close donor depth at 20nm we observe that the energy difference is approximately equal in magnitude to the TCAD potential at the nucleus. This is because the donor wave function has perturbed only slightly from the zero field ground state wave function. In contrast the energy difference for the donor depth at 40nm is much higher as the wave function deforms significantly from the ground state wave function towards the applied voltage.

If we compare the results obtained in our work including the effect of the interface barriers in addition to the electric field potential, with Kane's\cite{kane2} results wherein only the potential of a uniform electric field in the bulk was considered, we observe that the silicon oxide layer and the back gate exert a significant influence on the donor electron's ground state. Instead of the contact hyperfine coupling, $A(V)$, being independent of whether a positive or negative voltage is being applied at the $A$-gate as reported by Kane, we observe in Fig.~\ref{fig:fige4} that the interface regions in the silicon host geometry break this symmetry. 

Even without considering the influence of the interface regions, the effect of whether a positive or negative voltage is applied at the $A$-gate causes very different changes in the donor electron density. For a positive voltage the electron is bound to both the nucleus and the $A$-gate. In contrast, when a high enough negative voltage is applied so that the electron is no longer bound to the P nucleus, the electric field profile causes the electron to disperse in all directions away from the positive potential.

\subsection{Results obtained varying back gate depth and donor depth }

\begin{figure}
 \includegraphics[height=3in,width=2.5in,angle=-90]{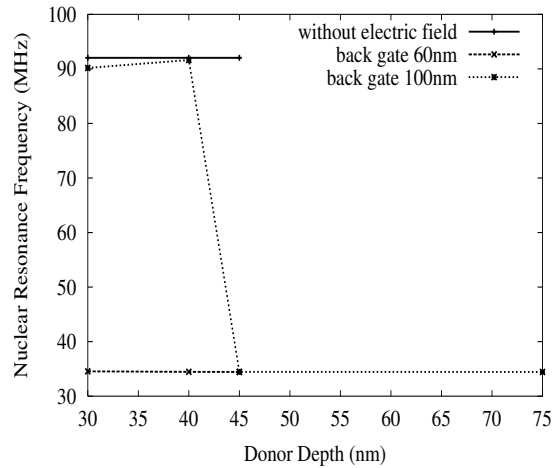} %,width=7cm,height=11cm,clip=,angle=270}
 \caption{\label{fig:fige9} Nuclear resonant frequency shifts of qubit $Q_1$ at varying donor depths with back gate depth at 60 and 100nm, and 1.0V at $A$-gate. }
 \end{figure}

To observe the effect that the back gate depth has on the donor electron wave function we repeated the calculation with a back gate depth at 100nm. Figure~\ref{fig:fige9} shows the comparison between nuclear resonant frequency shifts of the donor electron with the application of a voltage at the  $A$-gate with a close and far back gate. These calculations were performed with a close back gate at 60nm and a far back gate at 100nm, with a bias of 1.0V at the $A$-gate and donor depths ranging from 30 to 75nm.

With a closer back gate the electric field strength was higher within the Si wafer, and the donor electron wave function was perturbed greater, and so the frequency shift was more pronounced for donor depths with a close back gate. For donor depths close to the back gate the interface barrier effectively ``pushes'' the electron towards the $A$-gate. With the back gate at 100nm, the electric field strength is lower, and there is no substantial overlap of the donor electron wave function with the back gate barrier for donor depths of 30 and 40nm, so it is not as energetically favorable for the donor electron to perturb away from the back gate toward the $A$-gate.

\begin{figure}
 \includegraphics[height=3in,width=2.5in,angle=-90]{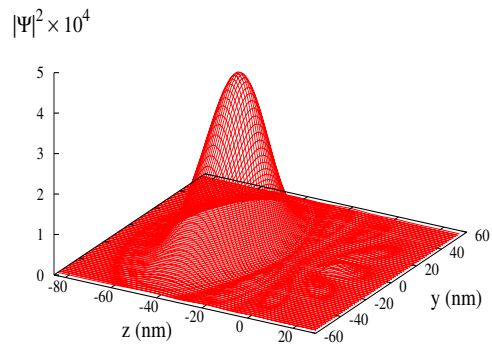} %,width=7cm,height=11cm,clip=,angle=270}
 \caption{\label{fig:fige9a} Ground state electron density in $yz$-plane for donor depth at 75nm with back gate depth of 100nm, and 1.0V at $A$-gate. }
 \end{figure}

Figure~\ref{fig:fige9a} shows the ground state wave function plotted in the $yz$-plane for a donor depth of 75nm and with a back gate depth of 100nm and a positive voltage of 1.0V at the $A$-gate. This plot demonstrates that even at a donor depth far from the $A$-gate, the ground state wave function distorts freely toward the $A$-gate because of the close proximity of the back gate, and the remoteness of the silicon oxide interface.

In Table~\ref{table2} we present a comparison of the difference in the ground state energy for the donor wave function without the electric field, and with a voltage of 1.0V applied to the $A$-gate, and the back gate at 100nm. Also reported in this table is the TCAD potential at the P nucleus for the varying donor depths. This table reflects the trend noted in Table~\ref{table1} that a significantly lower ground state energy is obtained for the deeper donor depths, where the electron density perturbs significantly away from the nucleus toward the applied voltage. 
\begin{center}
\begin{table}  [h!]
\caption{$ E_{1V} -E_{0}$ for a back gate depth of 100nm. \label{table2} }
\vspace{0.3cm}
\begin{tabular} {|l|l|l|l|l|l|} \hline
$Q_1$ Depth & TCAD Potential  & $ E_{1V} -E_{0} $  \\ 
 & at $Q_1$ (meV) & (meV)  \\  \hline
40nm & -67.25    &  -68.33  \\ \hline 
75nm& -26.00 & -36.57  \\ \hline
\end{tabular}
\end{table}
\end{center}

\section{\label{sec:five} Conclusions and implications for current fabrication technology and device modeling}

We believe that the results reported here using effective mass theory are quantitatively reasonable. It is reasonable to expect that the variation of the donor wave function with the experimental parameters calculated here using the smooth donor envelope function, would be similar to that of the true wave function.  

It is evident that the P donor electron wave function is sensitive to all experimental parameters studied in this paper. The donor wave function exhibits a fundamental change at crucial experimental parameters, where the electron wave function transforms from being only slightly perturbed from the zero field ground state, to being almost completely perturbed from the nucleus. These results highlight the significance of the influence of the silicon host geometry on the donor electron wave function. Ongoing work in our laboratory is focusing on verification of these results, using the full Bloch wave structure in our calculations. These results demonstrate the importance of the boundary conditions imposed by the interface regions, and the need to use a basis set which has the flexibility to meet the boundary conditions.
 
However, including the Bloch wave structure, the inter valley terms and the electric field and interface potentials is a challenging task. The results presented are quantitatively reasonable and provide a fast and reliable method which gives insight into the behavior of the P donor electron wave function under several different experimental conditions. To optimise the fabrication of these devices it is necessary to take into account the dependence of the donor electron wave function on all parameters varied in this paper: donor depth below the $A$-gate, back gate depth and voltage at the $A$-gate.

\begin{acknowledgments}
This work was supported by the Australian Research Council and the Australian Partnership for Advanced Computing National Facility. L.M. Kettle and H.-S. Goan would like to thank G.J. Milburn for valuable discussions relating to this work. H.-S. Goan would like to acknowledge support from a Hewlett-Packard Fellowship.
\end{acknowledgments}


\begin{thebibliography}{200}
\bibitem{kohn1} W. Kohn and J.M. Luttinger. Phys. Rev. \textbf{97}, 1721 (1955).
\bibitem{kohn2}  W. Kohn and J.M. Luttinger. Phys. Rev. \textbf{98}, 915 (1955).
\bibitem{kane} B.E. Kane. Nature \textbf{393} 133 (1998).
\bibitem{faulkner} R.A. Faulkner. Phys. Rev. \textbf{184}, 713 (1969).
\bibitem{koiller} B. Koiller, X. Hu, and S. Das Sarma. Phys. Rev. B \textbf{66}, 115201 (2002).
\bibitem{tahan} C. Tahan, M. Friesen and R. Joynt. Phys. Rev. B \textbf{66} 35314 (2002).
\bibitem{sham} L.J. Sham and M. Nakayama. Phys. Rev. B \textbf{20}, 734 (1979).
\bibitem{kane2} B.E. Kane. Fortschr. Phys. \textbf{48}, 1023 (2000).
\bibitem{larionov} A.A. Larionov, L.E. Fedichkin, A.A. Kokin and K.A. Valiev. Nanotechnology \textbf{11}, 392 (2000).
\bibitem{lloyd} C.J. Wellard, L.C.L. Hollenberg, C.I. Pakes. Nanotechnology \textbf{13}, 570 (2002).
\bibitem{Yu} P.Y. Yu and M. Cardona. \emph{Fundamentals of Semiconductors.} (Springer-Verlag, Berlin, 1996).
%\bibitem{koiller2}  B. Koiller, X. Hu, and S. Das Sarma. Phys. Rev. Lett. \textbf{88}, 27903 (2002).
\end{thebibliography}
\end{document}